# NON-PROFIT ORGANIZATIONS' NEED TO ADDRESS SECURITY FOR EFFECTIVE GOVERNMENT CONTRACTING

Lee E. Rice[1] and Syed (Shawon) M. Rahman, Ph.D.[2]

[1]School of Business and IT, Capella University, Minneapolis, MN, USA
`LRice6@CapellaUniversity.edu`

[2]Assistant Professor, Dept. of Computer Science, University of Hawaii-Hilo, HI USA
and
Adjunct Faculty, School of Business and IT, Capella University, Minneapolis, MN, USA
`SRahman@hawaii.edu`

## ABSTRACT

*The need for information security within small to mid-size companies is increasing. The risks of information security breach, data loss, and disaster are growing. The impact of IT outages and issues on the company are unacceptable to any size business and their clients. There are many ways to address the security for IT departments. The need to address risks of attacks as well as disasters is important to the IT security policies and procedures. The IT departments of small to medium companies have to address these security concerns within their budgets and other limited resources.Security planning, design, and employee training that is needed requires input and agreement from all levels of the company and management. This paper will discuss security needs and methods to implement them into a corporate infrastructure.*

## 1. INTRODUCTION

In developing a corporate security plan, it is important to understand the corporate structure and day-to-day business operations. The infrastructure of the Information technology department will be very important knowledge to have in order to build an effective security policy. The requirements for security are defined by the business and risk assessment. The corporate security policy is becoming more important to the overall company success and ability to attract customers.

In the development of a security policy all employees have a part in the implementation and success of the security policy. It is important for corporate officers to understand the importance of information security to the business. This will allow resources and money to be spent on this implementation and support. The Information Technology (IT) department must understand security and the best methods to utilize the corporate money and resources to get the best possible security policy put in place. The other employees need to follow the procedures and security guidelines for computer use to help prevent viruses, and unauthorized access to the systems through password compromise. These are important parts to the security policy in addition to the use of technology and proper tools to prevent and recovery from any possible attacks on the IT infrastructure.

In section two of this paper there will be a case study of a non-profit government contracting organization. This will show some of the challenges the IT structure needs to address. In section three of the paper common Information Attacks will be identified and possible solutions addressed. In the fourth section the security requirements for an organization of this type will be outlined and addressed. The fifth section will discuss security training for the employees and IT staff that would be helpful for any company. The sixth section will outline effective incident response for multiple issues and levels of problems for a company. The seventh section looks into the physical security needs for a company of this type. The eighth section addresses information assurance, the need to protect data and track data is very important in government contracting. In the final section will bring all of the sections together in a total security plan for a non-profit organization.

## 2. CASE STUDY: SEQUOIA ORGANIZATION

The Sequoia organization is a non-profit government contracting company. This company has developed a strong relationship with the federal government in the IT, environmental and health services sectors. This includes the expansion of the business to include prime government contracting and sub-contracting. The development of scientific equipment and solutions is part of the services this company provides through contracts. The security for this company is a very big part of the need based on the company direction and customer requirements.

The reasoning for choosing this company is based on familiarity with the corporate structure and the focus on security. It is important for a company to focus and care about security. This allows for a change and implementation of security changes easier than in a company that does not care about security. This company has multiple security systems and policies in place to secure the company from outside risk as well as internal risk. The protection and policies that are in place ensure the protection of corporate and client information and achieve the currently desired accreditations and certifications. The strong current posture that Sequoia has makes building a future security plan much easier.

### 2.1. Corporate Structure

Sequoia's business practices provides certain advantages for acquiring government contracts based on the different types and requirements for the agency looking for a contractor. The company has a standard corporate structure with a broad of trustees, Chief Executive Officer (CEO), Chief Financial Officer (CFO), Chief Technical Officer (CTO) and other standard corporate officers and positions. The internal Information Technology team report to the CTO which reports to the CFO. This corporate structure puts importance on the information technology for the company while still ensuring proper financial cost for the company.

We believe the current security posture for Sequoia is strong for a company of their size. The organization is less than 1000 employees. The role as a government contractor requires Sequoia to deploy multiple firewalls, internal permissions and security audits, as well as a high availability disaster recovery site. We believe in most organizations of this size the security need and requirements are not this extensive.

The IT infrastructure provides all of the internal corporate support as well as ensures IT solutions for the customers are properly configured and implemented properly. The security infrastructure oversees the implementation of all new corporate systems and ensures the policies and procedures for IT security are working properly. The Information Security Control Board is used to monitor, approve and internally audit all aspects of Information security (See figure 1).

### 2.2. Corporate Risk Assessment

The company has multiple checks and balances in place to mitigate risk and ensure corporate production despite outages or disasters. This includes strong firewall rules to protect the corporate data and systems. The internet sites are isolated from the rest of the network devices to minimize the exposure of the company and ensure data is protected while still providing information and access to the users on the internet. The internal network has security measures in place to protect the users from viruses, unauthorized access and data loss. The use of domain passwords, anti-virus software and backup and recovery methods ensure the company will continue to function properly for the users. In addition to these methods to mitigate risk, there are plans in place for disaster recovery of systems and the entire corporate office. This includes a business continuity site, off-site backup retention and secondary systems to maintain function and data integrity. There are ways to improve the disaster recovery process and standby systems. The company also needs to provide additional resources and policy for the future of the company and certain contracts the company will pursue. The corporate systems are in a virtual environment that allows for easy recovery and fault tolerance. These technologies help to develop a very good security plan for different types of risks or failures.

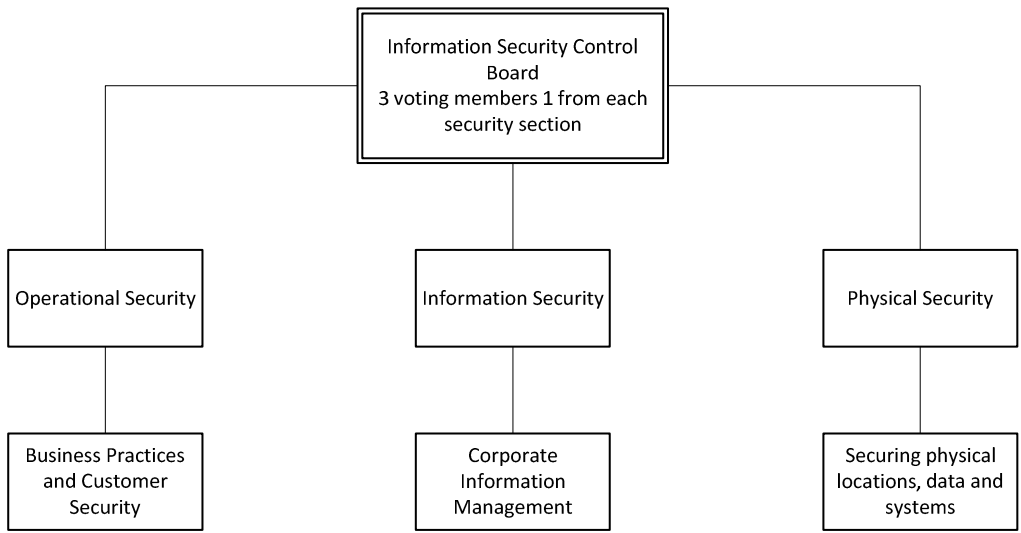

Figure 1: Security Architecture

This corporation has multiple networks and requirements for business objectives and customer missions. The different policies needed to protect the corporate data and customer data are very important to expand the business. In every network there are vulnerabilities and exposures that need to be considered and protected. This includes software and hardware risks within the infrastructure. In order to properly mitigate and control risks it is important to identify and address all areas risk can exist.

### 2.3. Network Infrastructure

Sequoia has a primary office in the Washington, DC metro area. This is the primary location for employees and systems for this company. In addition to this location there are a number of employees on customer sites and small satellite offices throughout the United States. There is also a location in rural Virginia that houses a super computer and disaster recovery location for Sequoia. The connection of the sites and users is achieved through multiple methods.

The network in the primary location has all the customer facing systems and corporate support infrastructure. This includes connections for the Internet, other corporate sites, and remote employees. The connections are achieved through Virtual Private Network connections, remote access connections and certain mobile applications. There are many external websites and applications that both customers and employees use to support the business. All of these applications require a strong and complex network infrastructure[1].

The primary network includes redundant circuits to the internet that support the customer information as well as the connection between the users and additional corporate sites. The network for the company consists of an external facing segment. This is isolated from the rest of the corporate network to protect the systems and data from web based attacks. In addition to the external network for web based application and external access. There is a DMZ network segment that provides limited access to the Internet and internal networks. This is where additional systems that support external connections and internal connections are kept. The network connectivity is limited by system and by ports to ensure protection for the data and internal network systems. The architecture for the network includes connections for servers and desktops on the internal network. This is for the main office, the satellite offices and the disaster recovery location. This provides domain, network and internet access for all systems (See Figure 2).

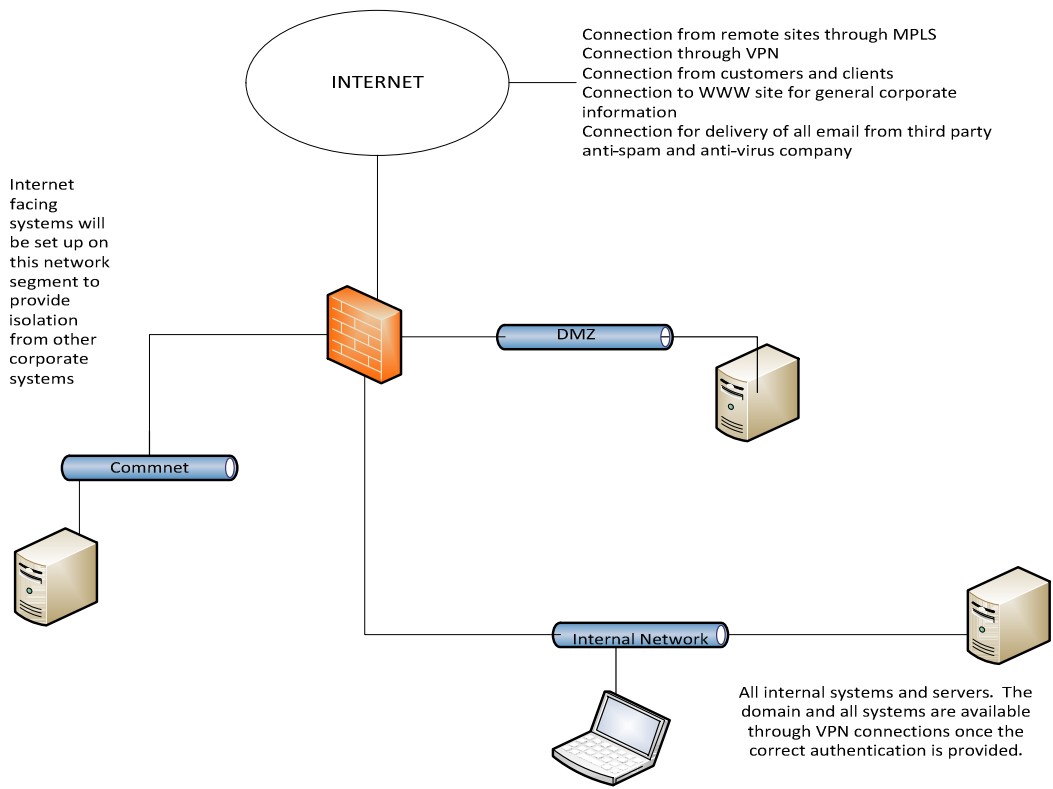

Figure 2: Basic Network Architecture

The basic infrastructure for the network is made up of standard network devices. This consists of switches, routers, and firewalls. The switches and routers are used to provide network connectivity for all users and systems. These systems also ensure traffic is being routed to the correct systems and networks. The firewalls are used within the infrastructure to determine what traffic is allowed to enter and exit the infrastructure. The network architecture provides access for employees and customers to use the systems of Sequoia.

### 2.4. Policy Structure

The security policies for Sequoia are very important to the business. In the government contracting business the security of the systems, network and data is very important to the customer. The Sequoia Corporation will need to build policies for network access. This will define what will be required for basic access to the network. The regulating of network access will help to understand what domain systems users have access to and what is restricted. This policy will ensure compliance with the US government standard for accessing data and information. This will verify the citizenship status for the users, to keep the data properly protected for the customer. In addition to basic network access policy the security policies need to include adding systems to the infrastructure, this policy will ensure notification and configuration of all systems on the network. The policy for system access from the internet will also be needed to protect the environment. This policy will document the business need for access and what ports are needed to be opened. Another policy that is needed to protect this infrastructure is the remote access policy. This will define what users have access to corporate systems from remote locations such as client sites and home. The policy will link user accounts with the separate authentication methods that will be used for the remote access. In the creation of security polices it is important to address potential risk and security breaches. The development of security policies will include disaster recovery as well. The DR policy is important because there are always potential issues that can cause outages or system failure. The ability to recover systems is vital to the sustainability of the company. The anti-spam policy will also help to secure the environment. This policy will govern the messaging systems to reduce messages and risks from unsolicited e-mails into the environment. All of these policies will be used to help secure the corporate infrastructure.

Table 1: Vulnerability Matrix

| Asset | Threat/Vulnerability | Impact | Mitigation |
|---|---|---|---|
| Physical Building | Theft | Varied(Minimum to Extreme) | Security patrol and surveillance systems |
| Physical Building | Fire | Varied | Fire Alarm and sprinkler systems |
| Physical Building | Natural disaster | Varied | Evacuation plan for employees |
| Data | Theft/compromise External | Extreme | Firewall, logging, audits, encryption and DMZ to protect data |
| Systems | External cyber attacks | Extreme | Anti-virus, port blocking, logging to assist in IP blocking |
| Data | Theft/compromise Internal | Major | Permissions, logging, audits and encryption |
| Systems | Accidental, malicious tampering- internal | Extreme | Physical Access limitations, local login rights heavily restricted |

### 2.5. System Vulnerabilities

The systems in the Sequoia infrastructure are vulnerable just like at any other corporation. The risk assessment should identify these risks to ensure they are considered and mitigated by the security policies that will be put in place. The vulnerabilities will range from external threats to internal threats (See Table 1). There are many threats to any company systems they cannot be addressed until they are identified properly.

Risks for Sequoia from the internet are based on the access to corporate systems from the Internet. This will open certain systems and data to access from the general public. It is important to understand what data and systems are able to be accessed from the Internet. The access from the internet can be anyone since the public site is on the internet. The risk for internet sites is based in unauthorized changes to the websites or data. In addition to the data changes on the website, the website can be used to break into other systems within the organization. The VPN solution for this company also creates vulnerabilities. This allows connections from the internet to access all corporate systems. These connections are tunnels through the internet that open corporate resources to authenticated users. The risk is that the tunnel will be compromised or unauthorized users gain access to the VPN tunnel. In addition to risks in the VPN tunnel and internet sites, there is the risk of gaining access through open ports and executing malware or other viruses. These risks from the internet are very important to mitigate in order to maintain network access and function.

The risks from inside a company can also cause issues with the systems and users. The internal risks are from accidental data loss, the risk of users deleting or incorrectly overwriting data can cause projects to be incorrect. The risk of viruses spreading and infecting systems can be a huge corporate risk including sending critical data to outside sources. In addition to these accidental or unknown risks there is the risk for malicious users attacking the company from within. This includes terminated employees and other reasons for employees to attack or steal data. There is also a risk for systemmalfunctions within the organization to cause network outages or congestion. The internal risks are usually more accidental or a malfunction than malicious in nature like the external risks.

The Sequoia Company has many risks that can affect the data and employees of the company. It is important to have a strong plan to address these risks and how the company will react to compromise of the company systems. The development of a security plan will include physical security plans, plans for incident response, implementation plan, and policies and procedures to govern the protection of information and systems. The development of a strong security plan begins with prioritizing of the risks in order to properly address them.

### 2.6. Prioritize Risks

The many risks that can impact Sequoia are similar to many other companies. However, every company prioritizes the risks differently. The business needs and customer needs will also dictate the prioritizing of risks within the organization. The IT budget is usually limited and the implementation of security can be a large cost for the company. Therefore it is important to prioritize the risks so that the important risks get mitigated properly within the budget and security plan.

In the organization the largest risk is data theft and unauthorized access to systems from outside the company. This risk will be addressed through the implementation of firewalls, network segment isolation (DMZ) and access control lists (ACL). These methods of mitigating network access risks will help the company to ensure data is not accessed from outside the company or web based systems are not exploited to compromise internal systems. In case of a compromise of a system or unauthorized access to the corporate network the IT team would isolate the IP address or system and take it off of the network until the system can be restored or the exploit fixed. The use of redundant systems and extra internet circuits will also be deployed to mitigate risk of denial of service attacks against the corporate network or systems. These network and Internet based attacks are a large portion of the risk for the Sequoia organization.

In addition to external network attacks, the company's next largest risk is natural disasters, and geographical attacks near the data center and corporate headquarters. These risks can cause massive loss of information, productivity and loss of the entire corporate infrastructure. In order to eliminate or lessen these risks the company must establish a secondary location to continue business operations. This location must be outside of the same power, network and physical location of the main data center at the headquarters for Sequoia. The need to have different power grid, network providers and establishing a good physical distance away from the primary location will help to mitigate any disasters or major outages of the utility providers to the primary location. In addition to network and proper power for the secondary site the site must provide systems to restore the production environment. This includes a copy or ability to restore corporate data to the remote site. The use of a disaster recovery site should be tested once every three months in order to understand the procedures and test the backup and recovery methods used for the critical systems. The analysis of the corporate systems to determine the critical applications and systems is also needed for this process. The use of redundant power supplies, power providers, network cards and network providers at the main location can also limit the vulnerabilities for Sequoia to the risk of attacks to the power or network providers. The need to recover quickly or mitigate disasters that affect the power, network or physical location is very important to maintain functionality for the company and customers.

The risk of internal data corruption, unauthorized access or malicious attacks is not a large concern for Sequoia but they are still a risk for the company. In order to mitigate these risks the Sequoia organization will use a strong access control system, employee screening, and deploy strong hiring and termination policies and procedures for all employees. The development of the permissions structure to ensure proper access to systems and files is very important to the protection of the network data from internal attacks. The exploiting of internal systems or users is also a concern that can be addressed through proper virus protection on all internal systems. In case of internal corruption or data lose the account responsible will be locked out of the network and any infected or causing computers will be disconnected from the network to prevent further compromise. These systems and users will be cleaned and evaluated to determine intent and extent of the damage. If a user is found to be purposefully compromising data they will be terminated as per the agreement that is signed upon employment with Sequoia. In addition to permissions the company will use employee education to help avoid internal compromise of data. If the users are educated on how to properly access data, understand the usual methods viruses use to access the systems and what to avoid when using the internet or corporate network they can be a great asset to the corporate security plan. The development of corporate auditing and monitoring will also help internal and external data loss or compromise.

These are the main risks facing Sequoia. The prioritization of these risks will determine how important they are within the security plan and the funds to protect the company from them. Some of the methods used to protect and mitigate certain risks will be helpful to reduce other risks as well. This ability to address multiple risks with a single implementation will be very god for Sequoia and the customers.

### 2.7. Hardware and Software Risks

The network and attacks areexcessive risks for any company. However, the hardware and software for the systems can also carry risk. The hardware in most cases is a single point of failure. If the hardware fails the data or application on that system is not available to the users and customers. This includes network appliances, servers and desktops for the users. These pieces of hardware are required to effectively perform the corporate tasks and provide services to the customers. In hardware there is also power to consider, the failure of power in the corporate office or the data center can cripple the company. The power and hardware are a great risk for the company that needs to be addressed.

The software a company runs can be vulnerable as well. This includes bugs in the software that can be exploited by malicious code or hackers. The permissions and access for the software can be exploited to gain access to data or change data to provide incorrect information for the company or the customers. The code within software can be accessed and edited to change the function of the software and create issues with the corporate functions. There are vulnerabilities in both over the counter purchased software and in house developed software. These can cause different code to be executed, data to be incorrectly accessed, and data leak or compromise.

There are many risks within a corporate infrastructure. These risks include the network, systems, hardware and software. The company must understand the risks for their company in order to properly mitigate and deal with risks.

In the development of a strong security plan it is important to also include the methods to be used to recover from an attack, loss of data or failure of systems. The Sequoia organization requires the ability to avoid outages and also quickly recover from any major failure or potential system loss. In addition to natural causes for system failure a major concern is acts of cyber terrorism. These attacks can cause the failure of systems, the stealing of information, as well as the compromising of data to mislead customers or employees. In the security planning process it is important to identify these risks, and determine the appropriate course of action for the company to mitigate or recover from these issues.

The definition for cyber terrorism is any act using computer or telecommunication devices to cause damage, create fear or further ideas or religious thoughts.[2] These are not restricted to major terrorist groups. Any act using the internet to destroy or disrupt can be seen as cyber terrorism. In the Sequoia organization the main cyber terrorism threats are the denial of service for customers accessing company resources, and the unauthorized access to company data or systems. These are the basic attack types the organization must guard against. There are many ways to achieve these attacks, the use of viruses and Trojan horses are used to attack a company and cause systems to fail or send out corporate information. There are denials of service attacks that seek to find systems and then send a lot of packets at that system to cause legitimate traffic to be denied or fail. There are bots and computer networks that continually send packets of data at the network for the company; this will cause systems or websites to stop responding properly which can cost a company money and productivity

## 3. ATTACKS FOR INFORMATION TECHNOLOGY

### 3.1. Types of Cyber Attacks

Although there have not been very many successful large scale attacks against the US government according to Wilson. There are many small attacks that cause havoc for companies. The main attacks for Sequoia would be a virus being sent via e-mail into the company that would cause damage to systems and/or send information back out of the company.[1] This risk is mitigated with strong anti-virus software and anti-spam filters for the e-mail systems.In addition to this attack the denial-of-service (DoS) attack against the websites and internet services could cause damage and loss of production for the company. The final major attack is the remote execution of code on the internet systems. This is usually done through known vulnerabilities in the software and operating systems hosting websites and data. In the corporate infrastructure there are threats for the physical power and network systems as well. It is important to protect the systems from both internet and system attacks as well as attacks on the infrastructure that provides connectivity and power to the corporate systems.

### 3.2. Source of Attacks

There are many sources for cyber terrorism. In the past cyber terrorism was suspected to be the acts of known terrorist groups furthering their agendas[1]. This is not always the case in the current corporate and Internet environment. The attacks can come from unknown sources trying to just cause havoc. Another source of cyber-attacks is parties out to damage the U.S. Government, which is a major customer of Sequoia. The final potential source of attacks is competitors of Sequoia, by causing failures or incorrect data to appear from the Sequoia organization this can discredit the organization or cause customers to make changes to the contract with Sequoia. The reliance on power and network resources opens the Sequoia organization to additional threats to consider. The potential attacks on the power and ISP can greatly impact the company's ability to properly function. Although the actual attacks are important to protect it is also important to consider where the attacks are coming from when trying to reduce the risk for Sequoia.

### 3.3. Tools for Prevention of Attacks

Table 2: Tools for preventing attacks

| Tool Name | Feature | URL for information |
|---|---|---|
| Symantec Anti-virus | Virus prevention, cleaning, and recovery | http://www.symantec.com/landing.jsp |
| Bluecoat Appliance | URL scanning, web filtering, malware prevention | http://bluecoat.com/solutions/business/web-filtering |
| MessageLabs | Third-party E-mail Anti-virus and anti-spam, content filtering | http://www.symantec.com/landing.jsp |
| Ironport Appliance | In networm e-mail anti-virus, anti-spam, content filtering | http://www.cisco.com/en/US/products/ps10154/index.html |
| Qualys Appliance | Vulnerability scanning and reporting covers many applications and operating systems. | http://www.qualys.com/products/qg_suite/vulnerability_management/ |

The prevention of these attacks is a very important part of a disaster recovery (DR) plan. Although recovering from disaster is the reason from the DR plan, the prevention of attacks or disasters is a better method for business continuity. The attacks outlined above call for specific tools to be used for the prevention of attacks. The denial of service attack is based on finding and overloading systems or IP addresses. This can be prevented with the deployment of a strong firewall in front of all of the corporate systems. The use of border systems to prevent malicious packets or traffic is vital to the proper function of the corporate systems. This device will provide rules for access and denying access to unauthorized systems or users.

The prevention of viruses can be done through multiple systems. The use of anti-virus software on all systems such as Symantec provides ways to prevent the execution and infection from virus files. This software will scan the system for potentially infected files or programs. This will ensure the virus does not spread or create more damage after the infection. In addition the anti-virus software on the systems the deployment of anti-spam and anti-virus filtering for e-mail is a great prevention of attacks or infection. There are services such as Message Labs as well as software and devices on the corporate network such as Ironport that can scan, filter and prevent messages that are spam or infected from reaching the corporate systems and users. The use of web filtering can also prevent viruses and other web based attacks from infiltrating the company. The tool for this type of filtering that can be used is a Bluecoat appliance. These measures will contribute to the prevention of viruses and damage to the systems.

The exploiting of operating systems and software is a very big part of the cyber terrorism and malicious attacks against a company. This is mitigated by consistent and accurate patching of the systems on the network. It is very important to stay up to date for the proper function of the systems. The use of appliances and software such as Qualys to monitor patching and vulnerabilities can greatly reduce the exposure and risk for a company to these types of attacks. These systems can ensure the proper function of the systems as well as keep them protected from vulnerabilities in the software or operating systems.

In addition to these tools and methods of prevention, the use of logging and audits are very important to the detection and prevention of attacks. This can show holes in the systems, and unauthorized changes to permissions and port access within the network. In addition to tracking excessive attempts to access systems and where these attempts are coming from, the proper identification of multiple failure attempts can help to ensure the attempts are never successful. In implementing a strong audit and logging structure it will help a company be proactive in the fight against attacks.

We think the preventive measures are strong permissions, restrictive firewall port access, and using DMZ's for internet facing systems. The isolation of internet facing systems is a great way to ensure data is not corrupted or used against the internal more secure systems. All of these features are deployed based on a strong security policy. However even the best defense has holes and weaknesses, so it is also important to have a strong disaster recovery and continuity plan to quickly recover and further mitigate any possible attacks against the company.

### 3.4. Strategies for prevention

The ability to quickly recover corrupt files or systems is important to any recovery plan. The use of redundant power supplies in the systems can help to recover from certain hardware failures of any systems. The Sequoia organization should have multiple ISP's providing Internet access in and out of the corporate infrastructure. This will help to ensure a network outage for the ISP or a specific connection the company can still function. The addition of multiple power sources can help to ensure systems function in case of a failure of a certain power vendor or circuit. The addition of strong UPS and even generators can also ensure the power function for the corporate systems in the even to power failure or attacks. The Sequoia organization would also need a separate facility to provide a backup location to preform data to day operations if a major attack or building failure occurred. These strategies will ensure the business function no matter the cyber-attack, natural disaster or other potential disaster or attack occurs.

In the Sequoia Organization there are many needs for security policies and procedures. This includes the employee use of systems and the physical building. The policy for personal security has to include the use of corporate systems in addition to the protection of company data. The personal security policy will compliment and reinforce the corporate security policy.

The security policy for personal use in the Sequoia Company will require multiple parts to address all of the aspects of personal security. There will be sections discussing the protection of data, use of corporate systems, and protection of personal information of the employee that the company requires. This includes the education of employees to better understand security and possible risks that can present themselves in the organization.

The training of employees to better understand security and how to avoid risk will be a great part of the initial employee training. This will include training to avoid opening e-mail messages from unknown source. This will help to reduce the spread of viruses throughout the organization. The security training will also need to cover the protection of passwords and badges or corporate identification from being compromised or lost. The education of employees to understand the corporate security policies and standards would be a strong start to ensuring compliance with the security posture of the company [3].

The security policy in addition to education will restrict access to certain areas of the network or Internet. This will help to protect data and systems from being compromised. [2] This includes monitoring and restricting all Internet access in order to keep spyware, malware and other virus access to a minimum from corporate systems. The policies and procedures will notify employees of the Internet usage expectations and the fact that monitoring will occur for all Internet traffic. The monitoring of network access will also be covered in the monitoring policy this will help employees know what data is available and what is not. The monitoring can help the company to understand what is being done on the network. This can help to identify inappropriate traffic or access attempts as well as protect the company from viruses and malware.

The personal security policy will also include the length and complexity of passwords for all employees. The passwords need to be difficult to crack, in order to ensure the safety of the network resources. The requirement of multiple authentication methods from external sources will also add to the personal security for the company. The security of laptop and mobile devices will also be needed to protect corporate data and personal information. The policy will require encryption for all laptops in order to reduce data loss if a laptop is lost or stolen. The use of mobile devices with corporate information will be limited to those devices the company can manage corporate information and remove the data remotely in case of lost or theft.

The personal security policies for Sequoia will address many aspects of security for the company and the employees. The corporate use policy to protect the company from inappropriate activity will be important to ensure protection and keep employees from compromising the company systems. The password and authentication policies will help to reduce the ability for unauthorized access or hacking of corporate passwords to compromise Sequoia data. The goals of any personal security policy will be to inform and educate the employees of expectations and monitoring being done within the corporate infrastructure.

The requirements for a corporate security policy will be important for the company to correctly identify and implement in order to properly secure the corporate assets. The requirements for security are based on the risks identified and the business need for the company. The need for security is high for the Sequoia organization. The role as a government contractor requires high levels of security for the systems and the data on these systems. The education of the corporate users can realty increase the security for the company and help to protect the data and systems. There are many areas that education and training will help the company. The documentation of requirements and identification of training methods and areas will be very important to the proper implementation and execution of the security for Sequoia.

## 4. SECURITY REQUIREMENTS

The risk assessment for Sequoia identified many areas that need to be addressed. This includes the network infrastructure, the hardware and software of the systems, and the requirements for the users to access the corporate systems. The network sites are all connected to the main office through an MPLS setup over the Internet. The network should be segmented to help protect internal systems and data at all times. This will include certain systems that require two forms ofauthentication through the ISA server to ensure accurate and secure communication. The DMZ will be used to provide support systems to the Internet facing applications as well as other systems that will have additional access from the Internet. The risk assessment and corporate direction will determine the requirements. In the Sequoia organization there are many factors that determine the requirements for the IT security. This includes certifications to attract business, corporate mobility and flexible working environment policies, the protection of data for users and clients, and keeping corporate systems functioning properly. These objectives must be achieved or assisted with the security policy.

The network infrastructure includes multiple sites that require being secured from unauthorized traffic and access. This is achieved with firewalls at each location and central administration of the rules for these firewalls. In addition to firewalls, there needs to be different network segments and access control lists to isolate the remote offices from the main location. This will ensure that a breach of security or virus has limited ability to infect other systems or the entire company. [6] The network will also isolate systems and applications that are accessed from the Internet from the rest of the corporate systems. These, Internet facing, systems are more vulnerable to attack or compromise because of the presence and exposure to all of the traffic from the Internet. The isolation of them will help to protect the users and sensitive data the company uses. The protection of the network also includes protection from the Internet traffic as well. This is level of security will be achieved through the addition of a Bluecoat proxy device. This device will manage the traffic to and from the internet and filter out content or websites that are deemed dangerous or inappropriate for corporate use. The better management of users' web access can greatly reduce the amount of spyware or malware the company is exposed to. The network infrastructure is the backbone of the corporate IT presence it is important that these systems are operating and online for the company to function.

The hardware and software of the systems within Sequoia must be protected as well in order to have a strong security policy. The protection of the hardware and software is in two parts. The server hardware and software, these are the parts that run the corporate applications and services that are provided to the employees and customers of Sequoia. The second part of the hardware and software protection is the employee desktop hardware and software. These systems frequently hold sensitive information and must be protected in case of loss or theft. The protection of the systems on the network will be important to ensure the data and applications are safe to use by the employees and customers.

The corporate server systems will be secured in a couple of ways.
1. **Secure location:** This is a physically secured location that will isolate the server and support systems from the general employee access. This will be managed by badge access including a pin for each authorized employee. The employment of two factors of authorization for access will create a more restrictive access process.
2. **Restrictive Logon Rights**: This will ensure only the authorized users can logon to or interact with the software on the servers. The implementation of a strong permissions structure will help to keep software safe and ensure unauthorized users cannot save or interact with these systems. When securing the hardware and software for server systems it is generally based on physical and logical limitation to access for these systems. This will protect the systems from malice as well as from accidental corruption or loss.

The protection of desktop hardware and software will also be needed to help the company remain secure. The protection of hardware in desktop or laptops is more difficult than the server hardware. [4] The building will be secure from unauthorized people; however in the Sequoia organization the use of laptops is standard to provide mobility. The securing of these systems will be done through the encryption of the hard drive. Although this does not provide prevention of theft or loss of the hardware, it does protect the information on the system from being accessed or compromised. The protection of software and software licenses for the desktops will be done through the limitation of access to install files. This will allow the IT department to ensure the company complies with the licensing agreements for the software and International Journal of Network Security & Its Applications (IJNSA), Vol.4, No.4, July 2012ensure the software is not used by unauthorized users. The management of the software and hardware for desktops is also important to the security of data and systems for Sequoia.

The security of all systems will also include virus protection. The Sequoia organization will use Symantec products to manage and prevent viruses from infecting the corporate systems. The Symantec products will also be used to protect the corporate system from spam and viruses within the email system. These products will help all systems to stay protected even in the event a virus is accidentally opened or downloaded. The use of preventive virus and spam protection is very good for all corporations.

The security for user access will be based on job function and preliminary screening of the employees. The network access for the users will be restricted based on the job function of the user. This will determine the permissions granted, account type and physical access within the building. The user access also will determine access to the VPN systems. This will allow users to access the corporate systems and network from a remote location. This will be limited to the Sequoia owned devices and require two form authentication, the user domain password, and a securid token. The use of two form authentication will be important to the protection of corporate systems while being accessed from the Internet. The use of job function based user controls will ensure that users do not have unneeded access to systems or data. This will protect the data from being corrupted or lost.

## 5. SECURITY TRAINING

The best defense for any company is the employees. The ability of employees to identify and prevent security risks is important and can greatly assist the company to stay protected from attacks and viruses[3]. The training of employees should cover multiple areas of the organization. The corporate in processing should include corporate security training and the departments should train based on the departmental needs.

The training of employees shows the commitment to security and ensures the employees are prepared to deal with events that may occur. The security training will help to inform the users of potential risks within the environment. The informed users will be able to make better decisions and be more aware of the security issues for the company. The knowledge for the users will help create stronger passwords that will not be easily compromised. The training will ensure better web usage on the corporate systems, as well as inform the users of what to avoid on websites. The security trained user will know to avoid emails from unknown senders and not to open unknown attachments from anyone. The security training will be important to informing the users and providing a strong corporate stance for what is expected.

The materials for security training are important as well. These materials need to cover the important areas for all employees and the IT department. The needed materials are a copy of the corporate security policy that should be reviewed by all employees on a yearly basis. This will reinforce the policy and address any changes that have taken place. The Arsenal Security Group[4] is an example of an organization that provides training to non IT staff on multiple corporate security issues. This includes identity theft, competitive security and many other training types. The materials that are needed are presentations, examples of different security violations and situations. The use of written documentation for password standards, expected web behavior for all employees, the network access and remote access expectations, and the physical building and laptop security guidelines will allow the company to ensure all users are informed of the expectations for the employees.[11] The use of web based training for security situations such as social engineering will be available for all employees to learn the desired corporate actions and how to avoid these scenarios. The ability to access the corporate security policy and procedures at all times will also help the employees to ensure the security is taken care of for the company. The situations to review and learn from will also help employees to workwithin the security guidelines for the company.

## 6. INCIDENT RESPONSE PLAN

The implementation of security requires multiple levels of coordination and planning. This includes how to respond when a risk is realized. There are many levels to response for incidents. From simply restoring a file to deploying the disaster recover site for an extended period of time. It is important for any company to have a plan for common scenarios and a team ready to enact that plan. In the Sequoia Organization the Corporate Information Management (CIM) team is used to perform this plan.

The incident response plan requires proper identification of the risk and the impact to the company. These are the major factors to determine the course of action. The implementation of these steps and plans will be determined by the operations manager. In the Sequoia plan constant uptime for all systems is not required in the event of a major event. The business has deemed one week to be sufficient for critical systems to be restored based on the current customer need and the budget for the continuity for Sequoia. In order to provide faster recovery Sequoia is implementing disk to disk backup and the use of more online systems for recovery. However this does not shorten the restore and recovery requirements at this time. This incident recovery system will be used for all outages or security based incidents.

Table 3: Incident Impact Analysis

| Incident | Impact | Course of Action |
| --- | --- | --- |
| File deleted or incorrect version saved | Minimal | Restore file as soon as possible |
| Virus infection | Varying | Remove system from the network and isolate all infected machine in order to clean and contain the infection of corporate systems. |
| Security Breach | Varying | Using monitoring and logs isolate the IP address responsible for the attack and deny network access. Monitor logging and determine compromised systems and data to determine what needs to be restored or cleaned. |
| System failure | Medium | Reboot system, restore service for individual system, and contact vendor support if needed. Sequoia DR not made for redundant systems or failover in this instance |
| Power Outage/network outage | Major | Depending on duration, anything less than 1 business day the plan calls for waiting out the outage and restore service in the Headquarters location. Over one business day critical systems begin restore and recovery process at the DR location. Required up time for critical services 1 week per security plan. |
| Natural or other disaster | Catastrophic | Immediately begin building and recovering information and systems in the DR location. This requires constant work and recovery of systems within 1 week or sooner. |

## 7. PHYSICAL SECURITY

In addition to the logical and network risks that were addressed in the previous section, there are physical risks for the Sequoia organization that need to be addressed. This includes all of the offices, rooms within the main buildings and the disaster recovery site. The need to provide measures to monitor, prevent access and track employee access is very important to the security of the company. The physical security also extends to the protection of the employees from many different threats. The security team and system needs to address both of these areas effectively.

The implementation of physical security will include badge readers, cameras and keyed locks for the headquarters and all remote sites. The badge readers will be for all common areas and the server rooms at these locations. It will include a picture of the employee on the badge to help ensure the proper person is using this badge. In addition to the badge readers there will be cameras in the building and outside the building to monitor for suspicious or malicious behavior. The use of keys will be for the office doors and specific rooms. This will allow the company to regulate the key distribution and usage. The keys will be signed for and tracked by the main security office at headquarters as well as a security office in each remote office if needed. All of the monitoring and badge access will be monitored and managed from the main security office in the headquarters building this will reduce the need for staff and multiple management points. In addition to these steps a security guard team will be used to regulate visitors in the headquarters building as well as patrol the area in and around the building for unauthorized access or suspicious behavior. The guards will be responsible for responding and investigating all door alarms that are trigger due to potential unauthorized access. The disaster recovery center will be monitored and security maintained by the main security office, to protect the assets and ensure physical security since this building will not house employees on a regular basis. This building will only be monitored through cameras and security systems since the use of guards would not be as effective for this DR site. These physical controls will help to protect the building and company assets.

In order to protect the employees the cameras will be used to monitor the parking areas and building to ensure unwanted people are not hanging around the corporate offices. In addition the building will have fire alarm and sprinkler systems to ensure the quick evacuation and salvaging of building resources in the event of a fire. These alarm systems will also be used for any need to evacuate the buildings. There will also be plans for lockdowns and shelter in place to prepare for any potential risk. The security team will develop and test these plans as well as employ company employees for each area to assist in the implementation of these plans.

The protection of employees is a primary goal of the Sequoia organization. It is important to protect the employees as well as the physical location for the data and systems. The implementation of this physical security plan will ensure notification of problems, investigation, and prevention of any risks to the employees or the buildings. It is important to be proactive with monitoring and surveillance of the grounds and buildings.

## 8. INFORMATION ASSURANCE

The Information Assurance policies and procedures will be used to protect the information and maintain the integrity of the data within the organization. These policies will be used to classify data and how it should be protected. The procedures will be used to implement the objectives of the policy.

The policy for information assurance will be based on the need for Sequoia to obtain and maintain FISMA certification. This is required for the government contracts the company has and is looking to obtain. It is important to protect all data but it is extremely vital to protect personal information of employees and customers. The use of data encryption is the main method of securing data. This includes SSL, VPN, email encryption, Secure File Transfer and database encryption. In the Sequoia organization these are the main methods of Information Assurance.

The implementation of information assurance is done to keep the data intact and verify it has not been tampered with once it is accessed. The use of a VPN system for access corporate information from outside the organization will require two forms of authentication (domain password and securid code) then all communication will be encrypted to protect the systems and data that is traveling across these connections. The encryption of the databases will protect all of the information within the database and require secure communication between the web servers, application server and all other connections to the database. This will keep the most vital corporate information secure from being accessed from outside of these methods. The use of SSL will secure all connections and traffic to and from the web servers within the organization. This will help to protect the information and systems that provide information and connections to the corporate systems. These methods will be defined within the information assurance policy. In order to provide access to personal or corporate sensitive data approval will be needed from the Information Security Board which will require these authentication and security measure be implemented.

The procedures for information assurance will be to submit requests to obtain the ability to access corporate systems from outside the network (VPN). The procedure will require a business justification as well as manager approval in order to limit the access and minimize potential risks. The implementation of systems that access or make available personal, financial or customer data will require approval, proper planning and periodic auditing of security risks. The protection of information will also require training for all of the users within the company. This will be specific to job function and data that is available to the users. The training will ensure the users understand how to protect data and beware of methods to illegally gain access to the data. The information assurance procedures will work to ensure the protection of data and the ability to verify the data is uncompromised when obtained or used.

## 9. SECURITY IMPLEMENTATION

The implementation of the overall security plan must be a complete company effort and objective. It is not just an IT function to implement and maintain a security plan. This will require support from the corporate executives as well as the dedication to the cause of all employees. There will be hurdles for the plan to be successful, however they need to be addressed and mitigated to provide strong corporate security.

The implementation of this security plan will be done in phases. This will include planning, which will require the information gathering, risk assessment and risk prioritization. These steps can be difficult for any company. They require the executives to outline the budget and goals of the security of the company. This can be difficult information to gather and get to be specific enough. It is important for the security team to ask questions and use multiple methods to gather the needed information to create the proper plan.

The planning and implementation after the information is obtained can be a challenge based on the budget and need to be functional within the environment. The implementation of a security plan after a network or infrastructure is established is very difficult. The business will still need to function properly however certain aspects of the network access and traffic will need to change in order to be secure. It is important to audit and monitor traffic, port access, and permissions access before locking down and implementing all of the security features. This will help to ensure productivity does not stop while security is implemented.

The biggest aspect of implementing anything is the education of the users. In the terms of implementing a security plan it is not as much about getting the users to follow specific guidelines, it is more about the change of habits and behavior to become more security minded. This can be difficult as in most cases security is more time consuming and causes more work for the users. The best way to implement security is gradually and with a balance for the users and security needs. This will ensure users do not reject it and buy in to the philosophy that the company is implementing.

The implementation of security can be very beneficial to Sequoia if it is done with protection of the business without interrupting the business getting done. This will need to be done over time with user testing and input to ensure tasks are not avoided or circumvented by the user base. In the implementation of security it is finding the balance between secure and functional that will provide the best method to keep the company information, people and systems protected.

## 10. CONCLUSION

The development and deployment of a security plan for any company can be a very difficult task. However, in the small to medium companies this can be even more difficult. The lack of resources, personnel and budget to implement a strong security position can leave the company vulnerable to attack, outage, and theft of information. It is important to security plan properly; this includes a strong risk assessment for the company. Once a risk assessment is completed the company as a whole including all departments and corporate officers need to prioritize the risks and commit to the security requirements and posture that is agreed upon. The implementation and need for security is different for every company. However, it is no longer possible to ignore information technology security in any company no matter the business or size of the company. The protection of information and assets in the Internet and information age is critical to the success and privacy for any company.

International Journal of Network Security & Its Applications (IJNSA), Vol.4, No.4, July 2012

## AUTHORS BIO

**Lee Rice** (MCSE)holds a Bachelor's of Science in Information Technology/Information Systems Security from University of Phoenix (2010), and an Associate Degree in Information Technologyfrom University of Phoenix (2008). Lee Rice's current interests includeinformation security, messaging, mobile computing, and virtualization. He is currently enrolled in a Master's degree program in Information Assurance and Security at Capella University.

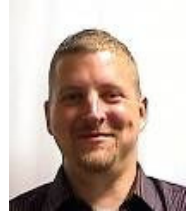

**Dr. Syed (Shawon) M. Rahman** is an assistant professor in the Department of Computer Science and Engineering at the University of Hawaii-Hilo and an adjunct faculty of information Technology, information assurance and security programat the Capella University. Dr. Rahman's research interests include information assurance and security,software engineering education, data visualization, web accessibility, and software testing and quality assurance. He has published more than 75 peer-reviewed papers. He is a member of many professional organizations including ACM, ASEE, ASQ, IEEE, and UPE.

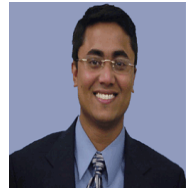